\documentclass[12pt]{article}

\usepackage{epsfig}

\newcommand{\ed}{\end{document}}
\newcommand{\be}{\begin{equation}}
\newcommand{\ee}{\end{equation}}
\newcommand{\ba}{\begin{eqnarray}}
\newcommand{\ea}{\end{eqnarray}}
\newcommand{\baz}{\begin{eqnarray*}}
\newcommand{\eaz}{\end{eqnarray*}}
\newcommand{\bb}{\begin{thebibliography}}
\newcommand{\eb}{\end{thebibliography}}

\begin{document}


\begin{center}
{\large{\bf Evidence for  unusual events in high energy  cosmic ray
                     interactions }}
\end{center}
A.K. Argynova $^{a}$, T.N. Kvochkina $^{b}$, A.A. Loktionov $^{a}$
\footnote{loction@satsun.sci.kz}
 N.I. Kochelev$^{c}$\footnote{kochelev@theor.jinr.ru}, A. V. Kropivnitskaya $^d$ ,
 A.A  Rostovtsev,$^d$\\
{\small\it
$^a$  Institute of Physics and Technology, Almaty, 480082, Kazakhstan\\
$^b$  Institute of Nuclear Physics, Almaty, 480082, Kazakhstan\\
$^c$ Joint Institute for Nuclear Research,
Dubna, Moscow region, 141980 Russia\\
$^d$ Institute for Theoretical and Experimental Physics, B.Cheremushkinskaya 25,
 117259, Moscow, Russia}

\vspace {0.2cm}

\begin{abstract}
 Anomalous events with  a large
 transverse momenta $P_T$ were found
for all secondary particles in
an analysis of experimental data on cosmic ray interactions at
$E_0 > 5\cdot 10^{13} eV$. In order  to clarify the  underlying
mechanism of these events, the intermittency
analysis of interactions with both large and usual transverse
momentum was performed. It is shown  that behavior of
intermittency indices  $\varphi (q)$ as a function of  rank factorial momentum
$q$  for such interactions is very peculiar. The indices show slight
growth for the events with large transverse momenta for all secondary
particles in contrast with the fast growth
for events with normal average transverse momenta for final
particles. The significant
difference in the behavior of intermittency indices  gives  strong
evidence   on the existence of the new class of hadron
interactions with large $P_T$ for all secondary particles in
the event. A possible interpretation of these anomalous events as a
 manifistation of  Color Glass Condensate
in  high energy hadron-hadron interactions  is given.
\end{abstract}

\newpage
\section{Introduction}

Investigation of the properties of quark-gluon interactions
at TEV region of energy is very important task, especially in the light
of future experiments at LHC.
One of the important goals of these experiments is clarification of
the structure of high density/temperature quark-gluon matter which can be
produced in heavy ion collisions.
At present, cosmic ray experiments give the unique information
on possible interesting effects in strong interaction of hadrons,
which could be carefully investigated at LHC.

A possible manifistation of Color Glass Condensate (CGC)
in high energy collisions is now widely under discussion (see recent
reviews \cite{CGC}).
Phenomena of CGC arise in the high parton density regime of
QCD where the typical virtualities of quarks and gluons are not very large
and thus their interaction is not very strong.
One of the most promising microscopical candidates for CGC is the
sphaleron state \cite{schrempp} which is closely related to the instanton
model of QCD vacuum ( see reviews \cite{reviewi})
The feature of the sphaleron driven multiparticle production is rather high multiplicity
in the events and large transverse momentum for all secondary particle.

In this Letter we will discuss a possible evidence for the formation of CGC in
high energy ($E_0 > 5\cdot 10^{13} eV)$ cosmic ray interactions with nuclei in the stratosphere.
An analysis of behavior of intermettency
indices for unusual type of the events, where
all the secondary particles  were produced with a  large
transverse momenta $P_T \sim 1  GeV/c$ \cite{dataour} is performed. An interesting
feature of these events is a large difference in $P_T$
distribution of the secondary particles compared to
their distribution in  the "usual" events, which have  $P_T\approx 0.2 GeV/c$.
Similar events were observed by JACCE \cite{JACCE}
and Concorde \cite{Concorde} Collaborations in  emulsion chambers.
In spite of the fact that the first data on existence of "anomalous"
events  were obtained about twenty years ago, the underlying
mechanism  of hadron-hadron interaction,  which is responsible  for
such events, is not understood so far.
A promising way in investigation of dynamics of multiparticle
production is the study of structure of the density fluctuations in the events
through calculation of so-called intermettency indices \cite{dremin}.
The values of the indices are very sensitive to the microscopical
mechanism of the hadron final state formation and, as we will show below,
are very different for  "anomalous" and "usual" cosmic rays  events.

\section{Experimental data}
In this Section we present a new analysis of the experimental
data,  obtained  in cosmic ray interactions at $E_0 >
5\cdot 10^{13}eV$, which were observed in $X$-ray film - emulsion chambers
exposed in the stratosphere at elevations $> 25 km$. Each
chamber contained target and gamma blocks. The target blocks were
gathered of many layers of heavy or light material interplead with
thin nuclear emulsions; gamma blocks contained 4 - 10 lead
plates, $5 mm$ thickness each with nuclear emulsions, and $X$-ray
films between them. The details of the construction of the
experimental chambers are described in \cite{dataour}.

For each event in the target block we determined
the number of charged secondary particles $(n_{ch})$, the number
of secondary $\gamma$ - quanta $(n_{\gamma})$, the angles of
emission of each secondary particle in the laboratory frame,
$\Theta_{ch}$, and, $\Theta_{\gamma}$ , an energy and a
transverse momentum of each over threshold $\gamma$ - quantum
$E_{\gamma}$ and $P_{T\gamma}$ (the threshold for the
electromagnetic cascade in $X$-ray films was $2 TeV$ and in
nuclear emulsion  - $0.05 TeV$). We estimate the deviation
of the accuracy for $P_{T\gamma}$ not greater than
$30\%$. The number of secondary charged particles and their angular
distribution provide an estimation of the
energy of the primary particles which produce interactions in the
target block of the chamber. The analysis of measured values of
transverse momentum of each photon in events with $\sum
E_\gamma > 10 TeV$ showed that 7 of them (index "$P_{T} = 1$")
have rather different properties in compare with an others.
Thus, the transverse momenta of most
photons in these 7 events were several times larger than the
average transverse momentum of secondary photons in the "usual"
type events with  $ < P_{T\gamma} > \sim 0.2 GeV/c$ (index "$P_{T}
= 0$").

The integral distribution of transverse momentum of all the secondary
photons is given in Fig.1 and can be described by the superposition of
two exponents:
\begin{equation}
   N_\gamma(> p_{t\gamma}) = A_1 exp(-p_{t\gamma} / p_{01}) +
    A_2 exp(-p_{t\gamma} / p_{02}),
\end{equation}
where $p_{01} \sim 0.2 GeV/c$  for the "usual" events and
 $p_{02} \sim 0.8 GeV/c$ for the "anomalous" ones. The main feature
of these 7 "anomalous" events is that most of the secondary photons have
transverse momentum $p_{t\gamma} \geq 0.5 GeV/c$.
The difference in characteristics between the two
classes of events in Fig.1 is large large compared to the errors in
the estimation of $E_\gamma$ and can not be explained by loss of the underthreshold
photons and statistical fluctuations.

\section{Analysis of the pseudo-rapidity distributions of the
secondary charged particles}
At first glance, both types of events
have a similar pattern. Thus, a comparison of $\chi^2$/NDF
fit with Gaussian distribution for events with higher mutiplicity
is given in table 1. The histograms of the pseudo-rapidity
distributions of the secondary charged particles for these events
are presented on Fig. 2. All events show approximately similar behavior.

\begin{center}

Table 1. Multiplicities and $\chi^2 / NDF$ of pseudo-rapidity  \\
\hspace {1 cm} distribution for large and small $P_T$ events  \\

\vspace {0.5cm}

\begin{tabular}{||p{1cm}||p{2cm}|p{2cm}||p{2cm}|p{2cm}||} \hline
\hline
 \multicolumn{1}{||c||}{ } & \multicolumn{2}{c||}{small $P_T$ events} &
 \multicolumn{2}{c||}{large$P_T$ events} \\
 \multicolumn{1}{||c||}{\bf N} & \multicolumn{1}{c|}{\bf NCH} &
 \multicolumn{1}{c||}{\bf {$\chi^2$}} &
 \multicolumn{1}{c|}{\bf NCH} & \multicolumn{1}{c||}{\bf {$\chi^2$}}  \\
  \hline \hline
  1 & 290 & 1.99 & 675 & 3.09 \\
  2 & 239 & 2.19 & 432 & 1.32 \\
  3 & 175 & 1.24 & 235 & 2.42 \\
  4 & 174 & 3.16 & 182 & 2.25 \\   \hline  \hline
\end{tabular}
\end{center}

\section{Analysis of the structure of correlations}

The analysis of the scaled factorial moments of the multiplicity distribution
to isolate the dynamical density fluctuations from the background of
statistical fluctuations was proposed in Ref.~\cite{bialas}:
\begin{equation}
       < F_q(\delta) > \quad = \frac{< n^{[q]} >}{< n >^q}
\end{equation}
where \ $< n^{[q]} > = < n_m (n_m - 1) \ldots (n_m - q + 1) >$.

In the presence of self-similar fluctuations of different sizes,
the dependence of the moment  $< F_q >$ on the size of the
phase-space bin  follows the  power law:
\begin{equation}
        < F_q > \quad = ( 1/\delta )^{\varphi_q}
\end{equation}
for $\delta  \rightarrow 0$. The positive constant $\varphi_q$  is the so-called
intermittency exponent.

The distribution (2) is not a smooth function, i.e. it contains
sharp spikes and holes between particles in the phase-space.
The observation of  the power law in a sufficiently  large  range
of  scales  $\delta$ indicates a self-similar fractal
structure of the short range particle density fluctuations.

In the opposite case of a smooth distribution (probability
density is continuous), the  factorial moments are
\begin{equation}
        < F_q > \quad \sim \quad Const.
\end{equation}

It should be mentioned that the high order moments resolve
the tail of the multiplicity distribution.
Thus they are very sensitive to the
density fluctuations at various scales of $\delta$ used in the analysis.
Therefore, the study of fluctuations at various values of
$\delta$, especially at small $\delta$ and large $q$ can improve  our
phenomenological  understanding  of  multiparticle   production
processes.

The particle multiplicity distribution is usually
studied for a sequence of phase space domains $\delta$
by consecutive subdivision of an initial region $\Delta$ in $M$
equal subdomains:

$$
     \delta = \Delta / M
$$
In every subdomain \ $m \quad (m = 1, ..., M) \quad n_m$ \  is the
multiplicity of that bin.

In order to improve  the statistical accuracy in the experimental estimation
of factorial moments, $F_q$'s of individual cells (1) are averaged
over events and over M cells ("vertical analysis"). The moments averaged vertically
({\it i.e.} over events) can be defined as a double average
\begin{equation}
< F_q > = \frac{1}{M} \sum\limits_m \frac{1}{N_{evt}} \sum\limits_{evt}
     \frac{< n_m^{[q]} >}{< n_m >^q},
\end{equation}
where \ $<n_m> = \frac{1}{N_{evt}} \sum\limits_{evt} n_m$ \
is the  multiplicity in bin $m$.

We  use a modified method of vertically
averaging \cite{method} in which moments are averaged at the starting point
of the original region $\Delta$ location
\begin{equation}
< F_q > = \frac{1}{M} \sum\limits_m \frac{1}{N_{step}}\sum\limits_{step}
     \frac{1}{N_{evt}} \sum\limits_{evt}\frac{< n_m^{[q]} >}{< n_m >^q},
\end{equation}
where  $N_{step}$ is the number of small $( step/\Delta \ll 1 )$
steps of the starting  position of the original region $\Delta$
in the area of pionization.
 The pseudo-rapidity $\eta = -
ln \ tg \Theta /2$ of the secondary particles charge is
used as the basic variable. The initial value of $\Delta$ is  4.0  and $M = 40$.
More detailed analysis has been performed with  $\Delta=3.5, 3.8 $, and $M = 35,38$,
respectively.

\section{The results}

The results  are presented in Table 2 and in Fig.3-5.
The values of $ln \ F_q$'s as a function of $-ln / \delta \eta$
for order $q=2$ to $q=8$, are plotted for small $p_t$ events in
Fig.3 and for  events with large $P_T$ in Fig.4.

For small $p_t$ events in Fig.3 the rise of
$ln \ F_q$ when increasing  $-ln \ \delta \eta$ takes place for
all $q$. For large $P_T$ events  (Fig.4) at all $q$ any
significant $\delta \eta$ dependence is absent.

 The slopes $\varphi_q$ in the region $0.1 \leq ln \ \delta \eta
\leq 1.0$ indicated by the solid  lines in the figures are given
in Table 2. Values of $\varphi_q$ are essentially larger
for the events with small $P_T$ compared to the large $P_T$ events.

The comparison of the slopes $\varphi_q$ for both types of
events is shown in Fig.5. For small $P_T$ events
the data are consistent with the intermittent behaviour (2), while for
"anomalous" events  the values of the slopes and their
changing as function of $q$  are
essentially smaller.

\vspace {0.5cm}

 Table 2. Slopes $\varphi_q$ fitted over
$0.1 \leq \delta \eta \leq 1.0$
         for large and small $P_T$ events

\vspace {0.5cm}

\begin{tabular}{p{2cm}p{6cm}p{4cm}}         \hline
   & small $P_T$ events & large $P_T$ events \\
 q            \\   \cline{2-3}
   & $\varphi_q$     &   $\varphi_q$   \\       \hline
 2 &  0.100 $\pm$ 0.004&  0.068 $\pm$ 0.005\\
 3 &  0.260 $\pm$ 0.014&  0.095 $\pm$ 0.010\\
 4 &  0.310 $\pm$ 0.027&  0.094 $\pm$ 0.016\\
 5 &  0.51  $\pm$ 0.05 &  0.08  $\pm$ 0.02 \\
 6 &  0.66  $\pm$ 0.06 &  0.10  $\pm$ 0.03 \\
 7 &  0.77  $\pm$ 0.09 &  0.11  $\pm$ 0.04 \\
 8 &  1.29  $\pm$ 0.11 &  0.13  $\pm$ 0.06 \\   \hline
\end{tabular}

\vspace {0.5cm}

\section{Possible explanation of anomalous events}

Self-similar particle-density fluctuations have been observed in
various reactions \cite{dremin}. Thus, the
approximate scaling in respect to decreasing of  domain size
was found. In spite of the strong experimental evidence,
the effect is still rather  far from  understanding based on
fundamental QCD theory. One of the possible contributions, which might give strong influence
on behavior of
 scaling moments, is coming from Bose-Einstein correlations. However, these correlation
should not be  related to some conventional static source \cite{bialas2}.

There is a clear indication on $P_T$ - dependence of intermittency in different
processes \cite{biswas}, \cite{ajinenko}. An analysis
for particles with $P_T$ smaller or larger than  $0.15 \div 0.3 GeV/c$
in the same events reveals a strong sensitivity to the value of transverse
momentum. The results show that the slopes $\varphi_q$ increase by a
factor $2\div 4$ for value of cut
$P_T=0.15 GeV $. A similar but not that strong effect is observed if
the $P_T$ cut is moved to $0.30 GeV/c$.

Our results for small $P_T$ events confirm the power-law behavior given by Eq.2.
On a contrary, for the large $P_T$ events, expression (3) looks as a
very promising candidate for the behavior of the intermittency exponents.

The results on $P_T$ distribution, presented in Fig. 1, and the results related to the
intermittency ( Fig. 3 - 5 and Tables 1, 2),
strongly indicate the existence of a new class of events with large $P_T$ for all  secondary
particles. We should mention that intermittency dependence
for the "usual" events
might be understood as a result of perturbative QCD evolution of
partonic cascade at high energy. In this case the independent
hadronization of the  quarks and gluons in partonic chain should lead to
a significant intermittency patterns~\cite{dremin}.
For the "anomalous" large $P_T$ events,  we found that the
behavior of the factorial moments is very different compared
to that of the "usual" events.
From our point of view this is very important
result. It seems  that in this case the perturbative
QCD can not describe the data and a new mechanism for secondary particle production
should be involved. One of the possible candidates for this mechanism is the
contribution from nonperturbative
partonic interactions at high energies, related to the Color Glass Condensate
formation at large parton density \cite{CGC}.
In this case the process of particle production in very high energy nucleus-nucleus
interactions induced by cosmic rays looks like simultaneous burst of all
interacted regions in the overlapped cosmic and target nuclei. If the CGC is identified with
the QCD sphaleron state \cite{schrempp},
which describes the particle production induced by strong non-perturbative
fluctuations of vacuum gluon fields, instantons \cite{reviewi}, than the
transverse momentum of produced particles should be roughly
$P_T(large)\approx 1/\rho_c\approx 0.6 GeV$, where $\rho_c\approx 0.3 fm $ is
the average size of instantons in QCD vacuum. This value of transverse momentum
is much larger than the value of  transverse momentum
$P_T(small)\approx 1/R_{conf}\approx 0.2 GeV$ ($R_{conf}\approx 1 fm$ is
the confinement size) of secondary particles in "usual" events.

\section{Conclusion}

In summary, a strong indication of the existence
of a new class of hadron-hadron interactions at high energies is
found. A possible  explanation of these interactions is a
manifistation of the Color Glass Condensate formation in the cosmic ray induced events.
It would be interesting to search for events with similar properties at
current (HERA, RHIC, TEVATRON) and future (LHC, TESLA) colliders.

\section*{Acknowledgements}
We are grateful to  A.E.Dorokhov for
useful discussion.
This work was partially  supported by grants RFBR-03-02-17291 and
 RFBR-04-02-16445.

\newpage
\begin{center}{\bf FIGURE CAPTIONS}\end{center}
\vspace{0.5cm}

\begin{description}
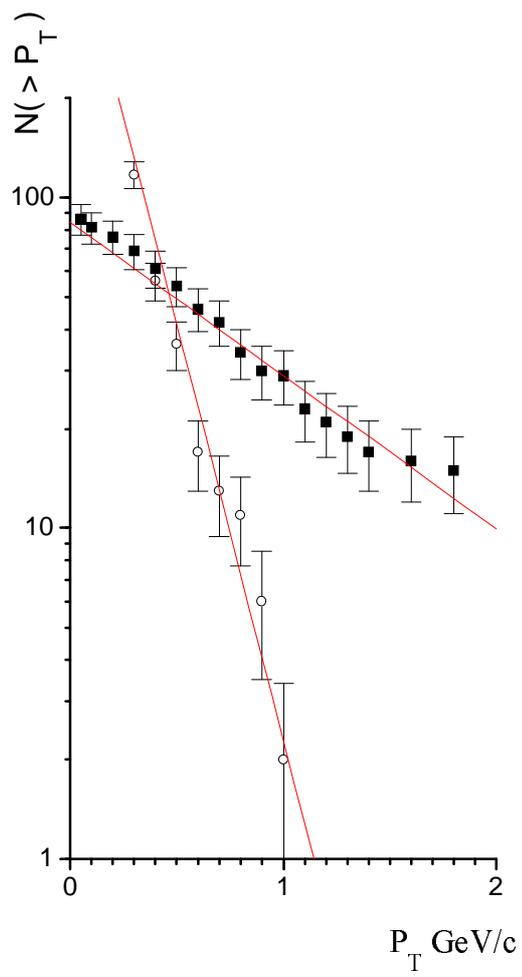
\item[Fig. 1.] $P_T$ distribution of secondary particles in two types
of events.           \\
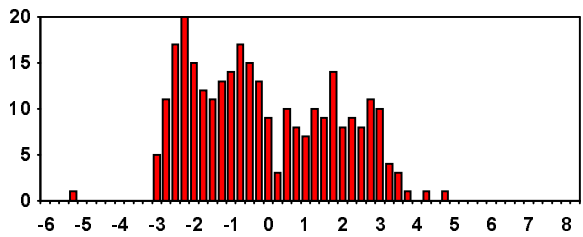
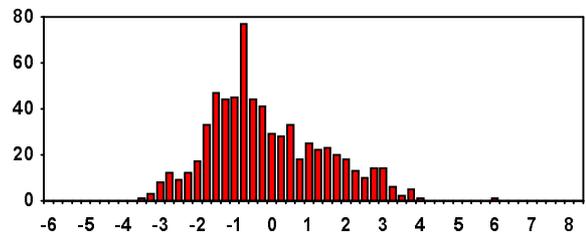
\item[Fig. 2.] Pseudo-rapidity distribution of events with  small
and large $P_T$.           \\
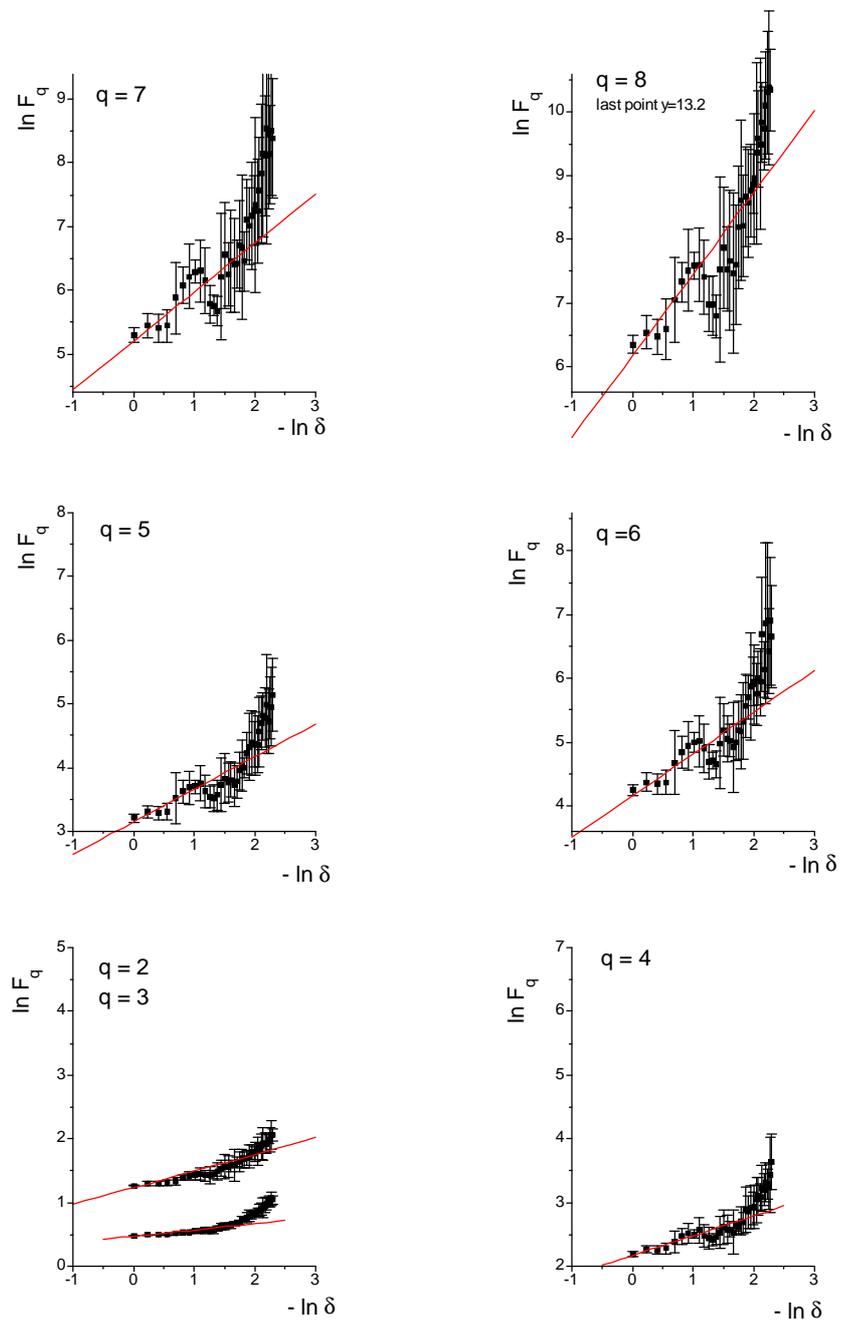
\item[Fig. 3.] $ln \ < F_q>$ as a function of $-ln \ \delta\eta$ for the group
of small $P_T$ events.     \\
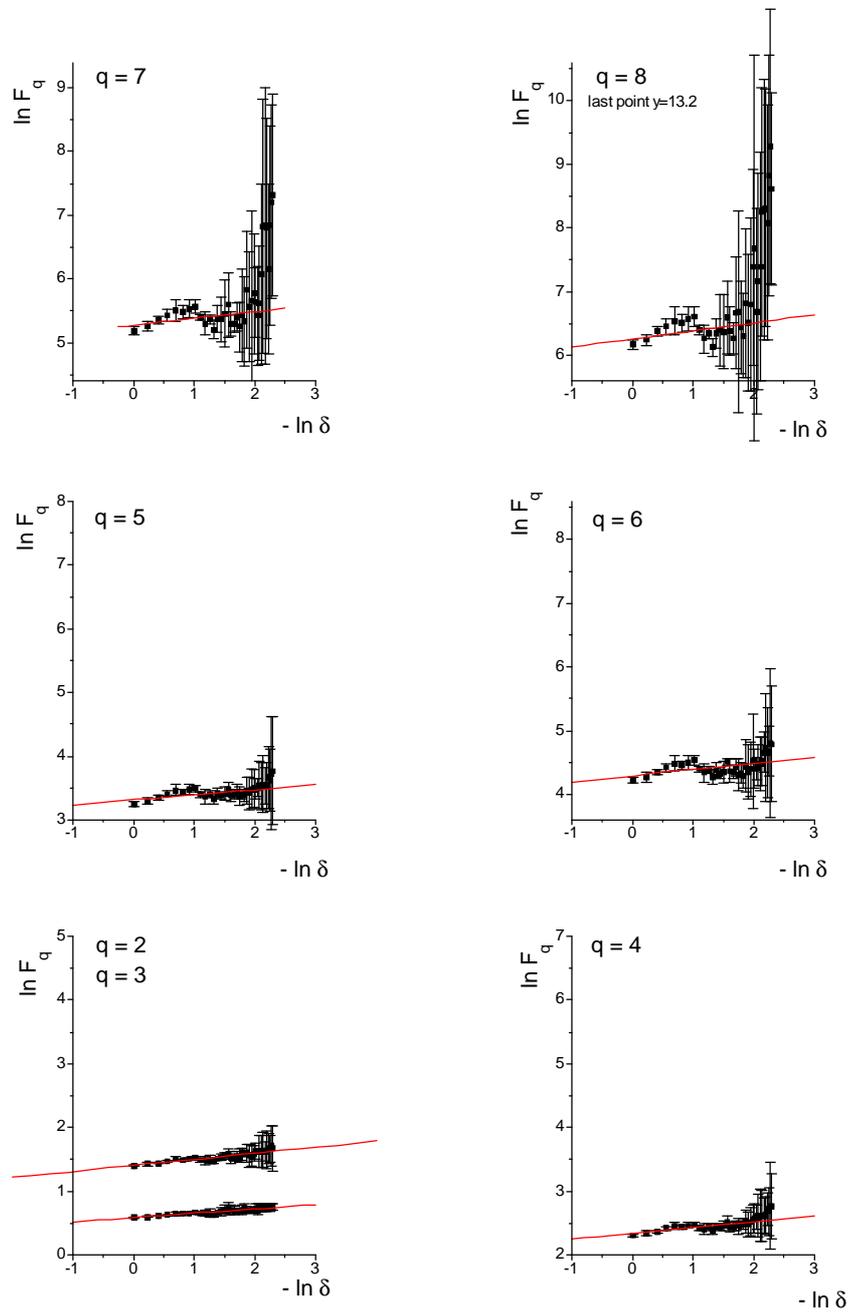
\item[Fig. 4.] $ln \ < F_q>$ as a function of $-ln \ \delta$ for the group
of large $P_T$ events.     \\
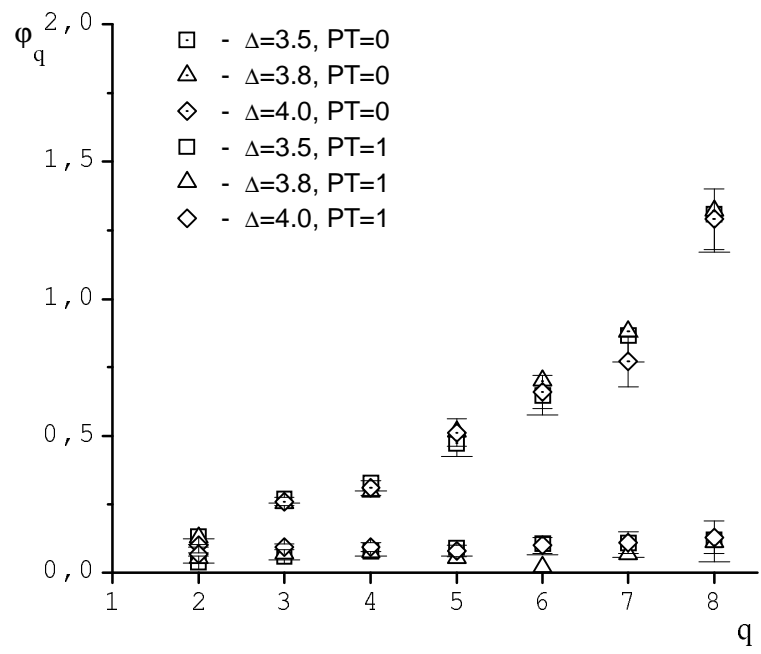
\item[Fig. 5.] The comparison of the slopes $\varphi_q$ as function of the
moment order $q$ for two types of events.
\end{description}

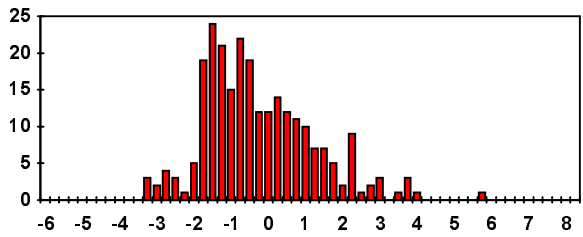
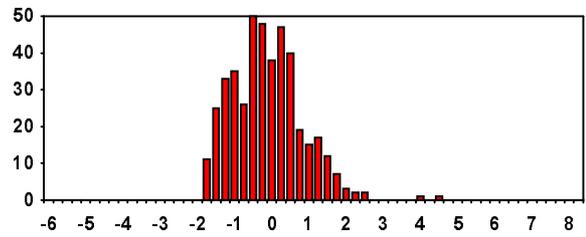
\end{document}